\documentclass[preprint, prb, amsmath, aps,showpacs]{revtex4}
\usepackage[dvips]{color}
\newcommand{\bea}{\begin{eqnarray} }
\newcommand{\eea}{\end{eqnarray}}
\newcommand{\bean}{\begin{eqnarray*}}
\newcommand{\eean}{\end{eqnarray*}}
\newcommand{\nn}{\nonumber \\}
\def\bfg#1{{\mbox{\boldmath $#1$}}}

\def\B{{\bf B}}

\def\E{{\bf E}}

\def\v{{\bf v}}

\def\zxi {{\bfg \xi}}

\def\btimes{~{\bf \times}~}
\def\bnabla{{\bf \nabla}}
\def\bcdot{~{\bf \cdot}~}
\newcommand{\lbs}{\left (}
\newcommand{\rbs}{\right )}
\newcommand{\lbm}{\left\lbrack}
\newcommand{\rbm}{\right\rbrack}

\def\av#1{\left\langle #1 \right\rangle}
\def\od#1,#2{\frac{d#1}{d#2}}
\def\odz#1,#2{\frac{d^2#1}{d{#2}^2}}
\def\pd#1,#2{\frac{\partial #1}{\partial #2}}
\def\pdz#1,#2{\frac{\partial^2 #1}{\partial {#2}^2}}
\def\eq#1{Eq.~(\ref{#1})}
\def\eqn#1{(\ref{#1})}
\usepackage{graphicx}
\usepackage{dcolumn}
\usepackage{bm}
\usepackage{epsfig}

\begin{document}

\bibliographystyle{unsrt}
%
%
\title{Peeling-off of the external kink modes at tokamak plasma edge}
%
%
%
\author{L. J. Zheng$^{a)}$ and M. Furukawa$^{b)}$ } 

\affiliation{$^{a)}$Institute for Fusion Studies, 
University of Texas at Austin, Austin, TX 78712
\\$^{b)}$Graduate School of Engineering, 
Tottori University, Tottori 680-8552, Japan}

\date{\today}

\begin{abstract}

It is pointed that there is a current jump between the edge plasma  inside the last closed magnetic surface 
and  the scrape-off layer and the current jump  can lead  the external kink modes to convert to the
tearing modes, due to the current interchange effects
  [L. J. Zheng and M. Furukawa, Phys. Plasmas 17, 052508 (2010)]. The magnetic reconnection
  in the presence of tearing modes  subsequently causes the tokamak edge plasma 
  to  be peeled off to link to the diverters. 
  In particular,  the peeling or
peeling-ballooning modes can become the ``peeling-off" modes in this sense.
    This phenomenon
  indicates that the tokamak edge confinement can be worse than the expectation based on the
 conventional kink mode picture.

\end{abstract}

\pacs{52.35.Py, 52.55.Fa, 52.55.Hc}

\maketitle

\section{Introduction}
 
The  H-mode confinement --- an operating mode with high energy confinement\cite{hmode}
--- has today  been adopted as a reference for  next generation tokamaks, especially for ITER.
 However, the H-mode confinement is often tied to
 the damaging edge localized modes (ELMs).\cite{hmode} There is a concern that 
ELMs can discharge particles and heat into the scrape-off layer and subsequently to the diverters.
The diverter plates can be potentially damaged by such a discharge.
This is particularly a concern for big devices like   ITER. 

This concern  has stimulated active researches in this field for clarifying the tokamak plasma edge instabilities,
in order to understand the ELMs. The most well-known theories are the peeling and 
the peeling-ballooning modes.\cite{wilson:1277,snyder:2037}
However, the peeling or peeling ballooning modes are of kink type. 
Without field line reconnection the plasmas inside
the last closed surface actually are not peeled off. 

The necessity to consider the tearing mode
excitation and the coupling of the scrape-off-layer current was first pointed out in 
Ref. \onlinecite{PhysRevLett.100.115001}. Apparently, to understand the ELMs one needs to take into consideration
the subtle feature of tokamak plasma edge, where the plasma on one side
is confined in the closed surfaces and  on the other side the plasma 
is linked to the diverters due to the open-field-line feature in the scrape-off layer. 
Otherwise, one cannot explain why there is not any  ELM-type of bursting at the internal transport barrier.
The development  of tearing modes can effectively connect the pedestal plasma to
the scrape-off layer. Taking into account this edge feature Ref. \onlinecite{PhysRevLett.100.115001} proposed a 
current-driving-mode theory for ELMs.  The magnetohydrodynamic (MHD) mode at plasma edge 
can be amplified due to the nonlinear coupling with scrape-off-layer current. This coupling can be 
a positive feed-back process and lead to the ELM bursting. The theory explains many
characteristic features of ELMs as observed at tokamak experiments, such as
 a sharp onset and initial fast growth of magnetic perturbations even when the underlying equilibrium is only marginally unstable for a MHD mode and also a quick quenching after the bursting peak. 
This work also points to the current driven modes --- tearing type --- as the ELM bursting explanation, although
the kink type of modes, such as the peeling ballooning modes, can be a trigger.  

In this paper we further explain how the external kink modes in tokamaks, such as the peeling ballooning modes,
 can become a trigger to the excitation of  tearing modes.
 We point out that there is a current  jump between the plasma edge inside the last closed surface and the scrape-off layer. When there is  a plasma perturbation at the edge, the currents on each side of the jump
  are carried over alternatively in the opposite direction to form a perturbed current sheet (see Fig. 1).
 This current sheet can lead to the excitation of tearing modes. 
 This mechanism reflects the extreme case of  
 the current interchange tearing modes as pointed out in Ref. \onlinecite{citm}, with the tokamak edge
 and scrape-off layer specialties being taken into consideration.
 Note that the drive to the current interchange tearing modes, as pointed out in Ref. \onlinecite{citm},
  is proportional to the current gradient. The current jump between the plasma edge and the scrape-off 
  layer make the drive at the edge to be dramatically  enhanced. As shown in the later on 
  analysis, the conversion of external kink modes
  to tearing modes at tokamak edge can therefore happen readily and cause the edge plasma to be peeled off. 
  Note that this  process may be positively fed back, as pointed out in Ref. \onlinecite{PhysRevLett.100.115001}.
  This phenomenon
  indicates that the tokamak edge confinement can be worse than the expectation based on the
 conventional kink mode picture.

We prove this peeling-off phenomenon by re-deriving the nonlinear tearing mode equation, which was originally
developed  by P. H. Rutherford.\cite{rutherford} Note that Ref. \onlinecite{rutherford} intended to consider 
the resistivity/current  gradient effects. However, it only took into consideration  the thermal conductivity effects related to the current gradient, without including the current convective effect as pointed
out in Ref. \onlinecite{citm}.  The current convective effect at the plasma edge can be very significant 
due to the  jump between the plasma edge and the scrape-off layer. This motivates us to 
examine  this issue. 

This paper is arranged as follows: Following to this introduction section, in Sec. II  the  Rutherford's equation will be rederived with the current jump between the plasma edge and scrape-off layer being taken into account; The results will be summarized and discussed in the last section, Sec. III.

\section{Rederivation of Rutherford's equation at the plasma edge}

In this section  we will rederive 
the  Rutherford's equation in Ref. \onlinecite{rutherford} 
to include the effects of the current jump between the plasma edge and scrape-off layer.
We first describe the Ohm's laws for the edge plasma and the scrape-off layer. For the edge plasma inside the last closed  surface  Ohm's law is
\bea
j_\| &=& \sigma E_\|,
\label{ohmp}
\eea
where $j$ is the current density, $E$ represents the electric field,  $\sigma$ is the conductivity, with
resistivity being $\eta=1/\sigma$, and subscript $\|$ denotes the parallel direction. In the scrape-off layer the field lines are connected to diverters at the both ends, indicated by $A$ and
$B$. The generalized Ohm's law in the scrape-off layer was derived in Ref. \onlinecite{solc}:
\bea
j_\|&=&\sigma_v E_\| -\gamma (0.85-\alpha)j_{SAT}\frac{T_B-T_A}{T_A},
\label{ohmv}
\eea
where
\bean
j_\| &=& j_{SAT} \hat j_\|,
\nn
\hat j_\|&=& -\gamma\bigg\{ \frac{e\phi_0}{T_A} +\lbs \kappa+0.85-\alpha \rbs\lbs \frac{T_B}{T_A}-1\rbs
\nn&&
+\ln\bigg[ \frac{1+\hat j}{(1-(T_B/T_A)^{1/2}\hat j_\|)^{T_B/T_A}}\bigg]\bigg\},
\label{hatj}
\nn
j_{SAT} &=&\frac1{2^{3/2}}enC_{s},
\nn
\sigma_v &=& \frac{e^2\lambda_{11} L_\|}{m_e}\lbm \int_A^B\frac{dl_\|}{n_e\tau_{ei}}\rbm^{-1},
\nn
\gamma &=& \frac{\hat\sigma T_A}{eL_\|J_{SAT}},
\nn
\kappa&=&\frac12\ln\lbs \frac{2m_i}{\pi m_e} \rbs =3.89.
\eean
Here, $e$ is the elementary charge, $m$ is the mass, $n$ is the density, $T$ denotes the temperature,
$\phi$ is the electric potential,  $\phi_0 = \phi_B-\phi_A$, 
$C_s$ is the sound speed, $\alpha  = \lambda_{12}/\lambda_{11}$, $\lambda_{11}$ and $\lambda_{12}$ 
are the Spitzer-Harm coefficients,\cite{spitzer} $\tau_{ei}$ is the electron-ion collisional time, 
$L_\|$ denotes the connection length between both ends $A$ and $B$,  $l_\|$ is the arc length 
along magnetic field line, subscripts
$e$ and $i$ represent respectively the electron and ion quantities, and subscripts $A$ and $B$ denote
quantities at the ends $A$ and $B$, respectively. Note here that the Ohm's laws in Eqs. \eqn{ohmp} and \eqn{ohmv} are given the moving frame. In the laboratory frame the electric field $\E$ needs to be replaced by
$\E +\v\btimes \B$. Here, we use the bold face to denote vectors, $\B$ denotes the magnetic field, and $\v$ is
the fluid velocity.

As Ref. \onlinecite{rutherford} we use the slab model in the $(x,y,z)$ space, with $x=0$
specifying  the rational surface and $z$ representing the longitudinal direction. The coordinate system is
shown in Fig. 1. The flux function $\psi$ and
the stream function $\varphi$ are introduced to represent the magnetic field $B_x= -\partial \psi/\partial y$,  $B_y=\partial \psi/\partial x$
and the velocity $v_x= -\partial \varphi/\partial y$,  $v_y=\partial \varphi/\partial x$. Here,  the
subscripts $(x,y,z)$ are introduced to denote the corresponding projections. We also introduce the displacement
$\zxi $, which is related to the velocity by $\partial \zxi/\partial t =\v$.

We consider the equilibrium with   magnetic shear, in which the poloidal magnetic magnetic field is represented by
 $B_y=B_y'x$. Here, prime
is used to denote the derivative with respect to $x$. The total magnetic flux can be written as\cite{rutherford}
\bea
\psi(x,y,t)&=& \psi_0(x) +\delta\psi (y,t),
\label{psi}
\eea
where $ \psi_0(x) = B_y'x^2/2 $ is the equilibrium value, $\delta \psi (y,t)= \delta \psi_1 (t)\cos ky$ is the perturbed
value, and $k$ is the poloidal wave number.  We use subscript $0$ to denote  the unperturbed quantities and
 ``$\delta$" to tag the perturbed quantities. Nevertheless, the subscript $0$ is dropped as soon as there is no ambiguity with the total quantities.
The purpose of this work is to prove that, if there is a free-boundary kink mode, it can be converted to the tearing modes
due the current jump from the plasma  region inside the last closed surface to the scrape-off layer. Therefore, we assume that there is a
 kink perturbation at the plasma edge  as follows
\bea
\xi  = \xi_0+ \xi_1 \cos ky.
\label{xi}
\eea
Here, $\xi_0$ is used to specify the distance 
between the last closed surface and the rational surface.  Note that at the plasma edge the magnetic shear
is very large, the distance 
between the last closed surface and the rational surface can be very small, so that one may assume $\xi_0\to 0$. We also note that the kink modes have different parity from
that of tearing modes. Although there is finite $\xi-\xi_0$ at the rational surface, the direct effect
of $(\xi-\xi_0)$ on $\delta \psi$ is negligible, since $\delta\psi \sim x (\xi-\xi_0)$. The effects of the displacement $(\xi-\xi_0)$
to be considered in this work is the formation of current sheet due to
the convective carrying-over of  equilibrium current. In difference from Ref. \onlinecite{rutherford},
in which  the $\xi-\xi_0$ turbulence effects on the tearing modes through the thermal conduction are considered,  
in this work we consider the convective effect on the formation of the current sheet.

As usual, we use the Ampere's law and the field diffusion equation
to construct the basic set of equations. The Ampere's law gives
\bea
\odz \delta\psi,x &=&\mu_0 \delta j_z,
\label{jz}
\eea
where $ \mu_0$ is the magnetic constant.

As for the field diffusion equation, we have to consider
separately the edge plasma region ($x\le 0$) inside the last closed surface and the scrape-off layer ($x>0$).  
We first consider the edge plasma region ($x\le 0$).
The derivation of the field diffusion equation is similar to that in Ref. \onlinecite{rutherford}.
Using Faraday's law one obtains $\delta E_z=\partial \delta\psi/\partial t$. Using this expression 
and the velocity representation with $\delta\varphi$, the curl operation of Ohm's law in \eq{ohmp} yields 
\bea
\pd  \delta\psi,t -\pd \delta\varphi,y B_y'x  &=&\delta(\eta j_z).
\label{pla0}
\eea
Here, as discussed previously, the $\v\times \B$ effect has been added in the Ohm's law \eq{ohmp}.
The perturbed quantity $\delta(\eta j_z)$ in \eq{pla0} contains both the local inductive ($\partial/\partial t$) and convective ($\v\bcdot\bnabla$) contributions due to the presence of the displacement $\xi$ in \eq{xi} (see Fig. 1). We exclude the inhomogeneity effects of the plasma resistivity both in the edge plasma 
region ($\eta$)
and in the scrape-off layer ($\eta_v$) from our consideration, since they are smaller than the effects
from the current jump between  the edge plasma  and the scrape-off layer. In consistence with
this we also ignore the inhomogeneity effects of 
other thermal quantities, such as $n$ and $T$. 
In the region where the edge plasma is not taken over by the scrape-off-layer plasma we then have
\bea
\delta(\eta j_z) &=& \eta\delta j_z.
\label{djp}
\eea
Instead, in the region where the edge plasma is replaced by the scrape-off-layer plasma
one has to include the convective effects due to the displacement $\xi$. This yields
\bea
\delta(\eta j_z) &=&\eta_v j_z+\gamma (0.85-\alpha)\eta_v j_{SAT}\frac{T_B-T_A}{T_A} -\eta j_{z0}
\nn
&=&  \eta_v\delta j_z -\Delta \hat E,
\label{djv}
\eea
where the electric field jump reads
\bean
\Delta\hat E \equiv  \eta j_{zp0} - \lbm  \eta_v j_{zv0} +\gamma (0.85-\alpha)\eta_v j_{SAT}\frac{T_B-T_A}{T_A} \rbm.
\eean
Here, $j_{zp0}$ and $j_{zv0}$ denote the equilibrium current densities  respectively in the plasma
edge and the scrape-off layer.  Using Eqs. \eqn{djp} and \eqn{djv}, the diffusion equation
in the edge plasma region ($x<0$), \eq{pla0}, can be 
expressed as
\bea
\pd \delta\psi,t -\pd \varphi,y B_y'x&=& H(\xi-x)  \eta \delta j_z+
H(x-\xi) \lbs \eta_v\delta j_z -\Delta\hat E  \rbs, 
\label{p2}
\eea
where, $H(x)$ is the Heaviside step function. Similarly, the diffusion equation in the scrape-off layer ($x>0$) can be
obtained as
\bea
\pd \delta\psi,t -\pd \varphi,y B_y'x &=& H(x-\xi)  \eta_v \delta j_z +
H(\xi-x) \lbs \eta\delta j_z +\Delta\hat E  \rbs.
\label{v2}
\eea
The current jump between the edge plasma and scrape-off layer and the inclusion of the
convective effects make the diffusion equations \eqn{p2} and \eqn{v2} become different from that in Ref. \onlinecite{rutherford}.

To proceed to derive the tearing mode equation, we still need  to consider
separately the edge plasma region ($x\le 0$) and the scrape-off layer ($x>0$).  We first treat  the edge plasma region ($x\le 0$). Dividing by $x$ and averaging over $y$ at constant $\psi$
 to eliminate the second term on the left,
equation \eqn{p2} becomes
\bea
\frac 1{\mu_0}\pdz \delta \psi ,x&=& \frac { \av{  \frac{\partial \delta \psi/\partial t}{(\psi-\delta \psi)^{1/2}} } 
 +\av{  \frac{H( x-\xi)\Delta \hat E}{(\psi-\delta \psi)^{1/2}} }  
 } 
 { \av{ \frac{H(\xi-x)  \eta+
H(x-\xi) \eta_v}{(\psi-\delta \psi)^{1/2}} }},
\label{p3}
\eea
where $\av{\cdots}= (k/2\pi)\int_0^{2\pi/k}\{\cdots\} dy$.  
Here, we have used \eq{jz} to express $\delta j_z$ on the left hand side and noted that $\delta j_z(\psi)$ is a function of $\psi$ 
only as required by the reduced vorticity equation $\B\bcdot \bnabla \delta j_z =0$,
 proved in Ref. \onlinecite{rutherford}.
Further integration over $x$ from $ -\infty\to 0$ of \eq{p3} yields
\bean
\left. \pd \delta \psi ,x\right |_{-\infty}^0&=& -\frac{\mu_0}{\sqrt{2B_y'}}\int_{-\infty}^0 \frac{d\psi}{(\psi-\delta \psi)^{1/2}}  \frac { \av{  \frac{\partial \delta \psi/\partial t}{(\psi-\delta \psi)^{1/2}} } 
 +\av{  \frac{H( x-\xi)\Delta \hat E}{(\psi-\delta \psi)^{1/2}} }  
 } 
 { \av{ \frac{H(\xi-x)  \eta+
H(x-\xi) \eta_v}{(\psi-\delta \psi)^{1/2}} }}.
\eean
Multiplying $\cos ky$ and averaging over $y$ this equation is reduced to  
\bea
\frac{\left. \pd \delta \psi_1 ,x\right |_{-\infty}^0}{\delta\psi_1} \delta\psi_1 &=& -\frac{2\mu_0}{\sqrt{2B_y'}}\pd \delta\psi_1,t 
\int_{-\infty}^0 d\psi \frac { \av{ \frac{\cos ky}{(\psi-\delta \psi)^{1/2}} } ^2
 } 
 { \av{ \frac{H(\xi-x)  \eta+
H(x-\xi) \eta_v}{(\psi-\delta \psi)^{1/2}} }}
 \nn
 &&- \frac{2\mu_0}{\sqrt{2B_y'}}
\int_{-\infty}^0 d\psi \frac { \av{ \frac{\cos ky}{(\psi-\delta \psi)^{1/2}} }\av{  \frac{H( x-\xi)\Delta\hat E}{(\psi-\delta \psi)^{1/2}} }  
 } 
 { \av{ \frac{H(\xi-x)  \eta+
H(x-\xi) \eta_v}{(\psi-\delta \psi)^{1/2}} }}.
\label{p4}
\eea
Introducing the dimensionless quantities $w=\psi/\delta \psi_1$, $ \Delta E= \mu_0 \Delta\hat E/(\eta B_y')$,
$\Delta'_-= {\left. \pd \delta \psi_1 ,x\right |_{-\infty}^0}/{\delta\psi_1}$, and the island width $x_T= 2\sqrt{\delta\psi_1/B_y'}$,
one obtains from \eq{p4}
\bea
\Delta'_- &=& \frac{\mu_0\sqrt{2}}{\eta}\pd x_T,t 
\int_{-1}^{+\infty} dw \frac { \av{ \frac{\cos ky}{(w-\cos ky)^{1/2}} } ^2
 } 
 { \av{ \frac{H(\xi-x) +
H(x-\xi) (\eta_v/ \eta)}{(w-\cos ky)^{1/2}} }}
 \nn
 &&+\frac{2 \sqrt 2}{  x_T}
\int_{-1}^{+\infty} dw \frac { \av{ \frac{\cos ky}{(w-\cos ky)^{1/2}} }\av{  \frac{H( x-\xi) \Delta E}{(w-\cos ky)^{1/2}} }  
 } 
 { \av{ \frac{H(\xi-x)  +
H(x-\xi) (\eta_v/\eta)}{(w-\cos ky)^{1/2}} }}.
\label{p5}
\eea
Similarly, in the scrape-off layer ($x>0$) one has
\bea
 \Delta'_+  &=& \frac{\mu_0\sqrt{2}}{\eta}\pd x_T,t
\int_{-1}^{+\infty} dw \frac { \av{ \frac{\cos ky}{(w-\cos ky)^{1/2}} } ^2
 } 
 { \av{ \frac{H(\xi-x)+
H(x-\xi)  (\hat\eta/\eta)}{(w-\cos ky)^{1/2}} }}
 \nn
 &&-\frac{2 \sqrt 2}{ x_T}
\int_{-1}^{+\infty} dw \frac { \av{ \frac{\cos ky}{(w-\cos ky)^{1/2}} }\av{  \frac{H( {\color{red}\xi-x}) \Delta E}{(w-\cos ky)^{1/2}} }
 } 
 { \av{ \frac{H(\xi-x)  +
H(x-\xi)  (\eta_v/\eta)}{(w-\cos ky)^{1/2}} }},
\label{v5}
\eea
where $ \Delta'_+={\left. \pd \delta \psi_1 ,x\right |^{+\infty}_0}/{\delta\psi_1}$.  Combining Eqs. \eqn{p5} 
and \eqn{v5} one finally obtains the tearing mode equation:
\bea
\Delta' &=& \frac{2\sqrt{2}\mu_0}{\eta}\pd x_T,t A_0
-\frac{2 \sqrt 2}{  x_T}A_c,
\label{total}
\eea
where $\Delta'=\Delta'_-+\Delta'_+$ and
\bean
 A_0&=&0.5\lbm \left. \int_{-1}^{+\infty} dw \frac { \av{ \frac{\cos ky}{(w-\cos ky)^{1/2}} } ^2
 } 
 { \av{ \frac{H(\xi-x) +
H(x-\xi) (\eta_v/ \eta)}{(w-\cos ky)^{1/2}} }}\right|_{x<0}
+
\left. \int_{-1}^{+\infty} dw \frac { \av{ \frac{\cos ky}{(w-\cos ky)^{1/2}} } ^2
 } 
 { \av{ \frac{H(\xi-x)+
H(x-\xi)  (\eta_v/\eta)}{(w-\cos ky)^{1/2}} }}\right|_{x>0}\rbm ,
 \nn
 A_c&=&- \left. \int_{-1}^{+\infty} dw \frac { \av{ \frac{\cos ky}{(w-\cos ky)^{1/2}} }\av{  \frac{H( x-\xi) \Delta E}{(w-\cos ky)^{1/2}} }  
 } 
 { \av{ \frac{H(\xi-x)  +
H(x-\xi) (\eta_v/\eta)}{(w-\cos ky)^{1/2}} }}\right |_{x<0}
+ \left. \int_{-1}^{+\infty} dw \frac { \av{ \frac{\cos ky}{(w-\cos ky)^{1/2}} }\av{  \frac{H( {\color{red}\xi-x}) \Delta E}{(w-\cos ky)^{1/2}} }
 } 
 { \av{ \frac{H(\xi-x)  +
H(x-\xi)  (\eta_v/\eta)}{(w-\cos ky)^{1/2}} }}\right|_{x>0}.
 \eean

Equation \eqn{total} is the modified Rutherford equation with the convective effects being taken into account
at the plasma edge, where there is a current jump. Note that in \eq{total} $\Delta'$ can be obtained from the outer solution,
 $A_0$ specifies the inductive contribution, and $A_c$ the convective contribution. Letting $A_c=0$
(i.e., $\Delta E=0$)  and $\eta=\eta_v$, equation \eqn{total} reduces to the usual Rutherford equation given in
Ref. \onlinecite{rutherford}. From Fig. 1 one can see that in the region for $H(x-\xi) =1$ and $x<0$ one usually has
 $\cos ky<0$; and in the region for $H(\xi-1) =1$ and $x>0$ one usually has $\cos ky>0$. Therefore, one usually
 has $A_c>0$. This shows that the convective contribution from the current jump  is generally a driving term for tearing modes.

Using the Ampere's law one can get the ordering estimate: $ \Delta E\sim {\cal O}(1)$. Noting that the second
term on the right hand side of \eq{total} is inversely proportional to the island width $x_T$, the convective driving contribution
can be very large. In the case with the current varying smoothly without a steep jump the convective driving term 
is proportional
to the displacement $\xi_1$ as shown in Ref.  \onlinecite{citm}. In the current case the current jump significantly enlarges the convective driving
effects in \eq{total}. Note that  the kink mode has a different parity from that of the tearing mode.
However, the inclusion of the current convective effects causes the two types of modes to become coupled.
This makes the kink mode is prone to convert to the current interchange tearing modes at the plasma edge.

To show the magnitudes and parameter dependences, we numerically compute the two parameters
 $A_0$ and $A_c$
 using the NAG (Numerical Algorithms Group) mathematical libraries, especially the subroutine D01APF.
We consider the  case with $\xi_0\to 0$. Figure 2 shows the dependence of $A_0$ on the resistivity ratio $\eta_v/\eta$. The displacement $\xi_1$
is used as a parameter in this figure, which is normalized by $x_T$. Figure 3 shows the dependence of $A_c$ on
the electric field jump $\Delta E$, with the resistivity ratio $\eta_v/\eta$ as parameter. Figure 4 shows the dependence of $A_c$ on the normalized displacement $\xi_1$ with
the resistivity ratio $\eta_v/\eta$ and the electric field jump $\Delta E$ as parameters. The calculations
show that the dominant contributions come from the current inside the magnetic island. 
From the parameter scans in these figures one can see that $A_c$ is of order unity. Equation \eqn{total}
shows that the convective contribution can be very big as compared to $\Delta'$.
This indicates that the perturbations of kink type at the plasma tend to convert to  the  tearing modes,
due to the current jump between the edge plasma and the scrape-off layer.

\section{Conclusions and discussion}

The release of thermal energy by tokamak plasma kink modes  has been widely studied in this field. In this paper we show that
the kink modes can carry over the equilibrium current and leads to the formation of the current sheet at the singular layer. Due to the vast difference between the equilibrium currents
in the edge plasma and the scrape-off layer, the current sheet can induce 
the tearing modes. This is an extreme case of the
so-called
current interchange tearing modes at the plasma edge
as pointed out in Ref. \onlinecite{citm}, with the tokamak edge
 and scrape-off layer specialties being taken into consideration.
Due to the current jump  between the edge plasma and the scrape-off layer, the driving effects for current interchange tearing modes at the plasma edge can be very big. Practically, any kink
perturbations on the plasma edge are potentially induce the tearing modes. The direct consequence
of the excitation of the current interchange tearing mode at the plasma edge is that 
the confined plasma inside the closed magnetic surfaces
can be peeled off   to the scrape-off layer and then to the diverters. As  an example, the peeling or
peeling-ballooning modes can become the ``peeling-off" modes in this sense.

What is more, Ref. \onlinecite{0029-5515-44-10-003} points out that the pumping out of the confined plasma in the closed
surfaces to the scrape-off layer can enhance the scrape-off-layer current, especially because the plasma edge
usually carries the negative charges, while the diverter sheets are  excessive in the positive charges.  The  the scrape-off-layer current can further drive the tearing modes and causes the positive feedback process. Therefore, the current work can help to explain further the edge localized modes in the H-mode confinement. 

Note that there is a similarity between the edge localized modes and the tokamak major disruptions. In the edge localized
mode case the scrape-off layer current is excited; while in the disruption case the halo current is induced. 
Both are explosive nonlinear processes and involve plasma and wall interaction.  
One is in a small scale; and the other is in a large scale. Peeling off
the confined plasma in the closed surfaces to the scrape-off layer or wall due to the current interchange tearing
modes at the plasma edge may also help to explain the disruption, especially the generation of the halo current
and its feedback. 

In passing, we note that  the current work has not  included the neoclassical tearing modes,\cite{ntm1,ntm2} 
although in principle the current interchange can include the interchange of the bootstrap current. 
This will be investigated in the future.

In conclusion, the possible excitation of current interchange tearing modes at the plasma edge  due to the current jump
 indicates that the tokamak edge confinement can be  worse  than the expectation based on the pressure driven (or kink) instabilities alone.

This research is supported by U. S. Department of Energy, Office of Fusion Energy Science and   by JSPS KAKENHI Grant No. 23760805.


\newpage

Figure captions

Fig. 1: The coordinate system for analyzing the current interchange effects. The axis $z$ points 
out of the paper. The perturbed current directions are indicated. The edge plasma locates at the $x<0$
region, while the scrape-off layer at the $x>0$ region. The plasma displacement $\xi$ is plotted by
the dashed curve with $\xi_0=0$ assumed. 

Fig. 2: The parameter $A_0$ versus the resistivity ratio $\eta_v/\eta$ with the displacement $\xi_1$ 
 as  parameter.

Fig. 3: The parameter $A_c$ versus the electric field jump  $\Delta E$, with the resistivity ratio $\eta_v/\eta$ 
 as  parameter. The normalized displacement $\xi_1=1$ is assumed. 
 
 Fig. 4: The parameter $A_c$ versus the normalized displacement $\xi_1$, with the resistivity ratio $\eta_v/\eta$ 
 and the electric field jump  $\Delta E$
 as  parameters.

\newpage

\begin{figure}[htp]
\centering
\includegraphics[width=150mm]{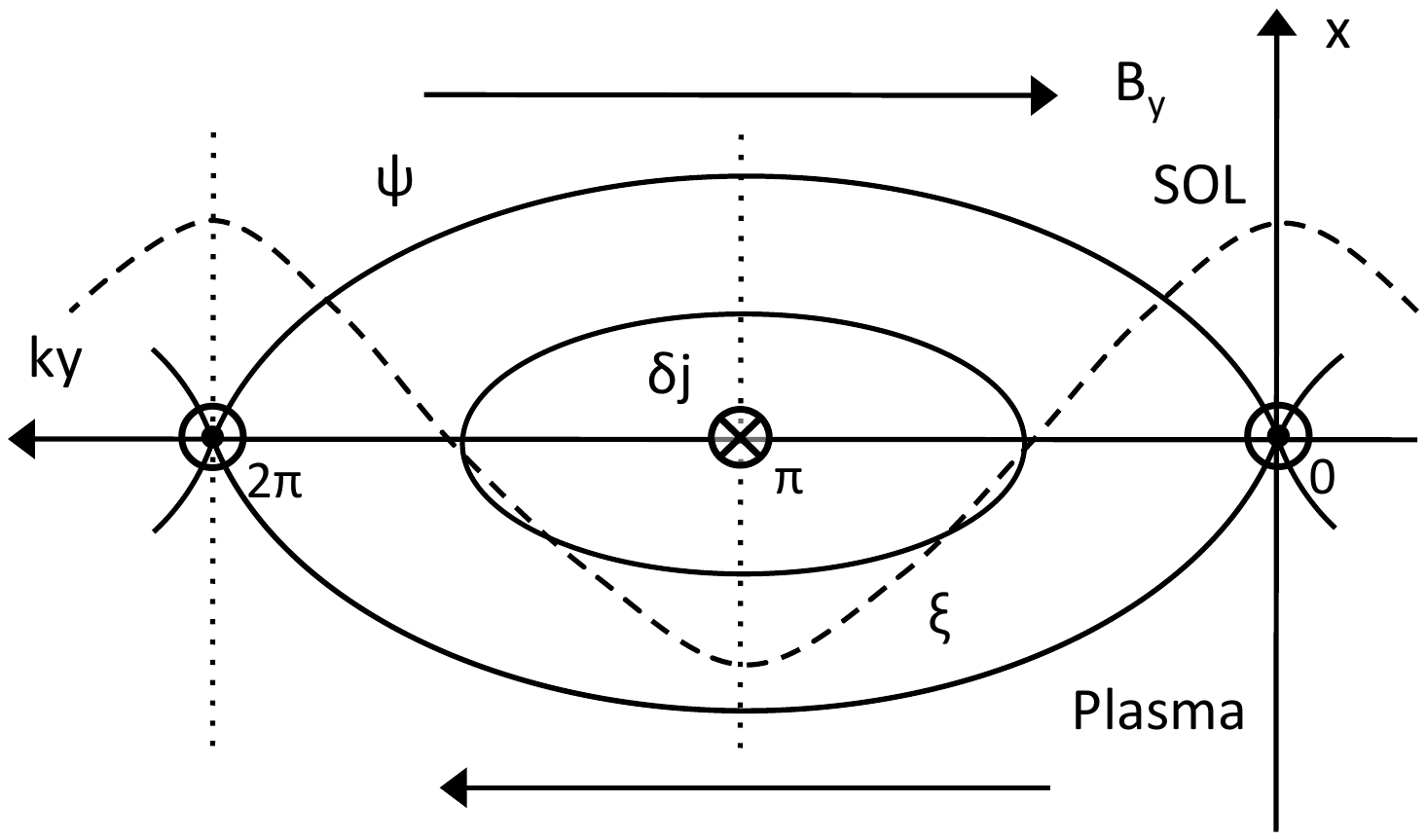}
\end{figure}

\centerline{  Fig.~~ 1}

\newpage

\begin{figure}[htp]
\centering
\includegraphics[width=150mm]{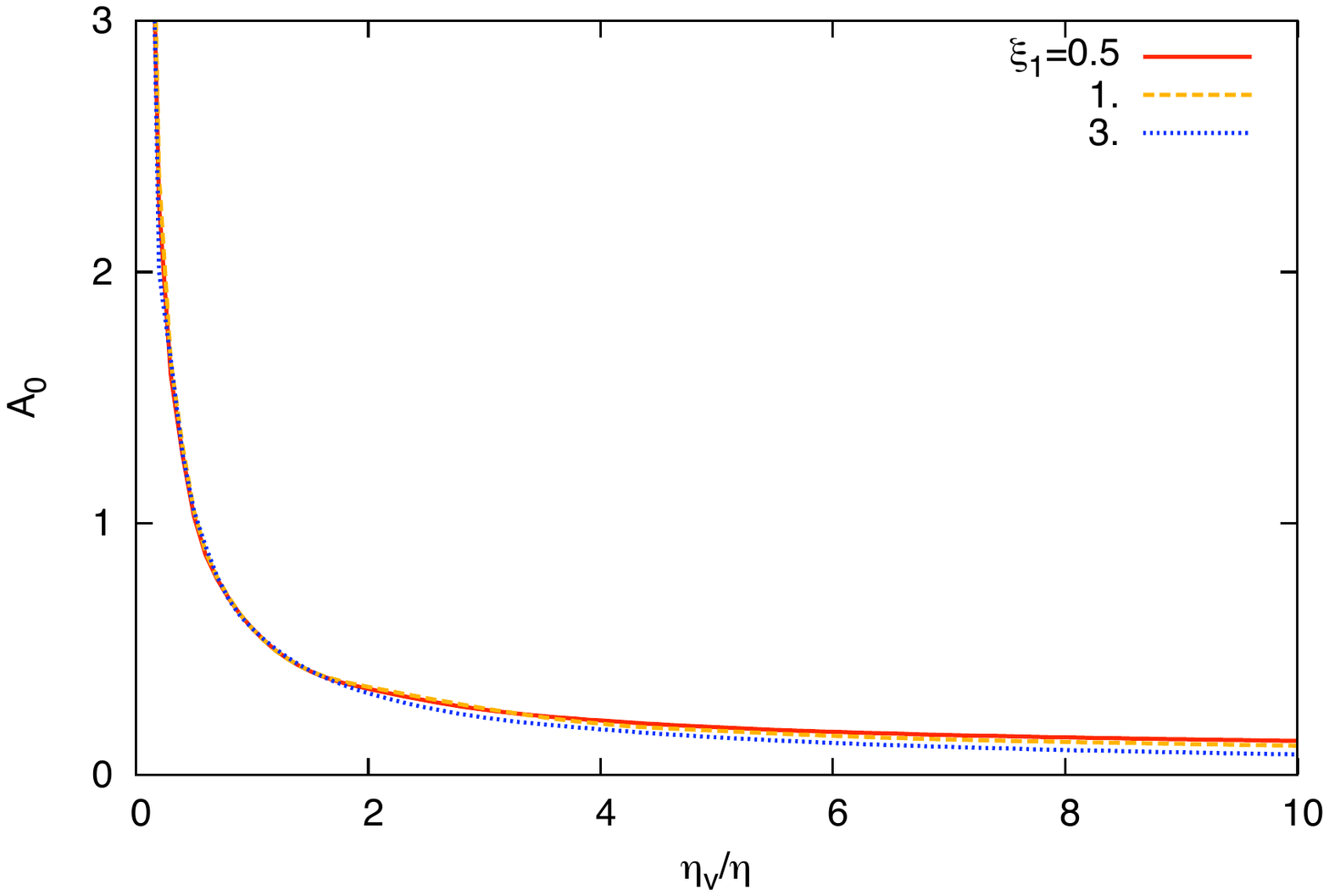}
\end{figure}

\centerline{  Fig.~~ 2}

\newpage

\begin{figure}[htp]
\centering
\includegraphics[width=150mm]{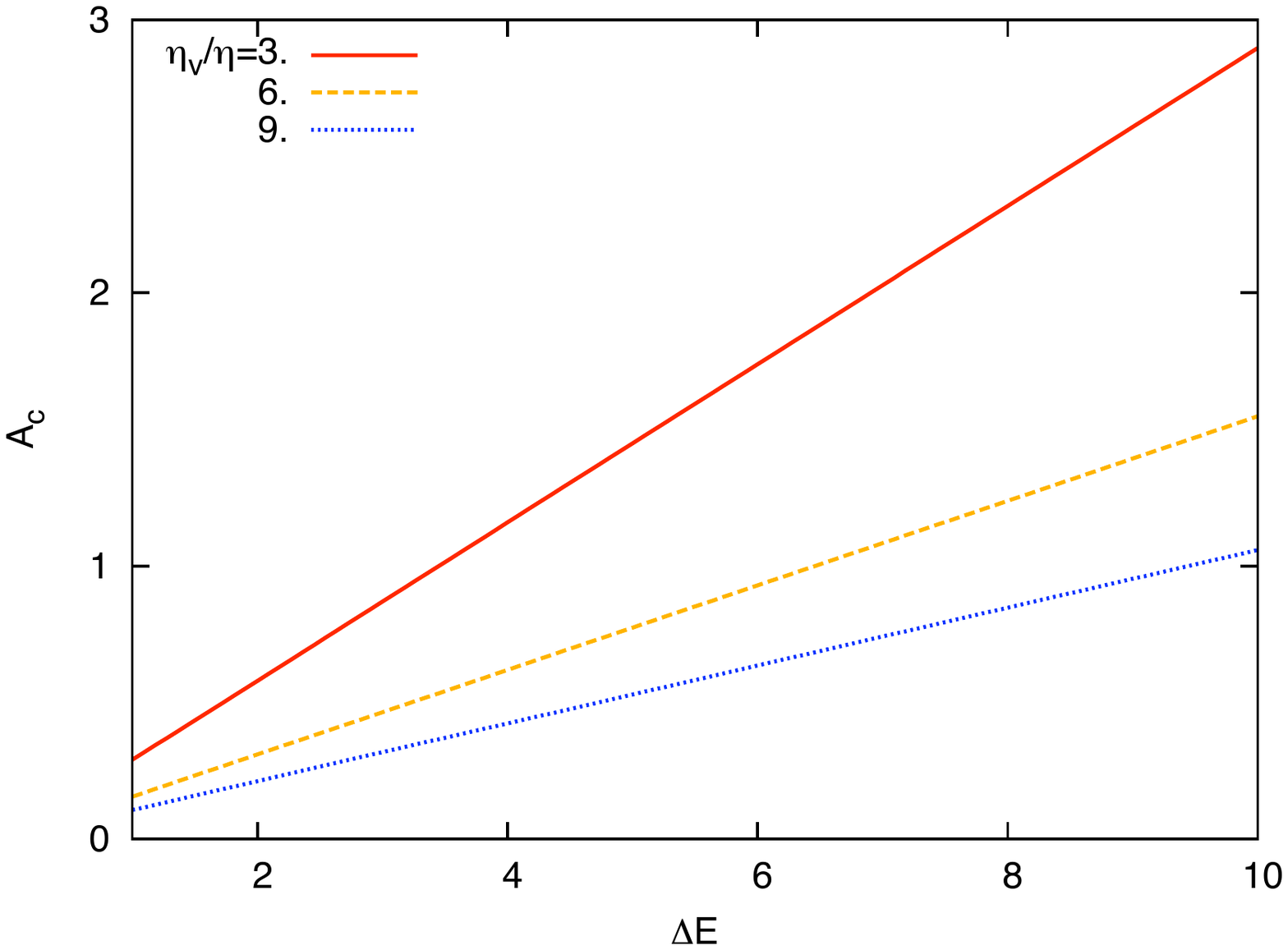}
\end{figure}

\centerline{  Fig.~~ 3}

\newpage

\begin{figure}[htp]
\centering
\includegraphics[width=150mm]{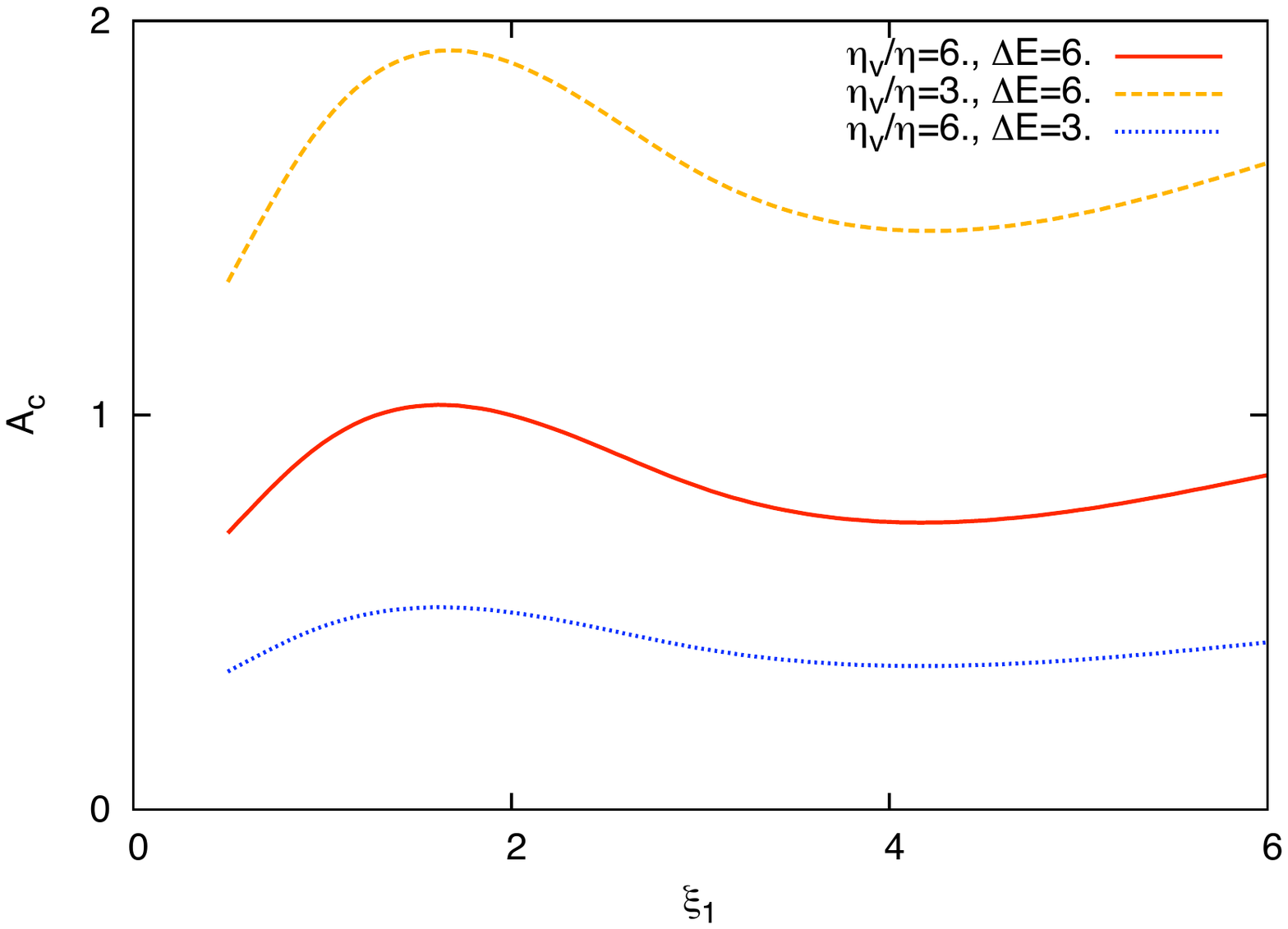}
\end{figure}

\centerline{  Fig.~~ 4}

\end{document}